\documentclass[10pt, conference, compsocconf]{IEEEtran}
\ifCLASSINFOpdf
  \usepackage{longtable}
  \usepackage{amsmath}
  \usepackage{color}
  \usepackage[pdftex]{graphicx}
  \usepackage{epsfig}
  \usepackage{epstopdf}
  \usepackage{graphicx}
\else
\fi
\begin{document}
\title{A GPU-accelerated Algorithm for Distinct Discriminant Canonical Correlation Network}


\author{\IEEEauthorblockN{Kai Liu, Lei Gao, Ling Guan}
\IEEEauthorblockA{\textsuperscript{} Department of Electrical and Computer Engineering, Toronto Metropolitan University, Toronto, Canada\\
Email: 09liukai08@gmail.com, iegaolei@gmail.com, lguan@ee.ryerson.ca}
}


%Telephone: (800) 555--1212, Fax: (888) 555--1212}


\maketitle

\begin{abstract}
Currently, deep neural networks (DNNs)-based models have drawn enormous attention and have been utilized to different domains widely. However, due to the data-driven nature, the DNN models may generate unsatisfying performance on the small scale data sets. To address this problem, a distinct discriminant canonical correlation network (DDCCANet) is proposed to generate the deep-level feature representation, producing improved performance on image classification. However, the DDCCANet model was originally implemented on a CPU with computing time on par with state-of-the-art DNN models running on GPUs. In this paper, a GPU-based accelerated algorithm is proposed to further optimize the DDCCANet algorithm. As a result, not only is the performance of DDCCANet guaranteed, but also greatly shortens the calculation time, making the model more applicable in real tasks. To demonstrate the effectiveness of the proposed accelerated algorithm, we conduct experiments on three database with different scales (the ORL face database, ETH--80 database and Calthe256 database). Experimental results validate the superiority of the proposed accelerated algorithm on given examples.
\end{abstract}


% \fi
\IEEEpeerreviewmaketitle

\section{Introduction}
In recent years, the deep neural networks (DNNs)-based algorithms have gained wide attention and have been utilized in different applications, such as visual computing, natural language processing, video analysis, among others [1-3]. In essence, DNNs-based models are able to generate deep-level feature representation by extracting the abstract semantics of the input data sets with a deep cascade network structure, leading to improved performance on various tasks. However, since the DNNs-based deep-level feature representations are learned from multiple cascade layers, it is necessary to collect enormous samples to make sure that the parameters in DNNs-based architectures can be tuned successfully. As a result, the DNN models may generate unsatisfying performance on the small scale data sets. In order to handle this problem, the multi-view representation method [4] is presented to balance the data scale and deep-level feature representation.\\\indent In [5], a multi-view representation algorithm,  canonical correlation analysis network (CCANet), is proposed. CCANet is capable of gaining more comprehensive information from multiple views, producing enhanced results on image classification. Nonetheless, due to its unsupervised nature, CCANet cannot explore enough discriminant representations. To deal with this issue, a discriminative canonical correlation network (DCCNet) [6] is introduced by employing within-class and between-class scatter matrices jointly. Nevertheless, it is acknowledged that the scatter matrix is only able to explore the discriminant information from single data set instead of multiple data sets. Therefore, a distinct discriminant canonical correlation analysis network (DDCCANet) [7] is proposed for image classification. Different from DCCNet, the correlation matrix is utilized in DDCCANet to explore the discriminant representations across multiple data sets, leading to improved performance on image classification. In addition, the presented DDCCANet is based on the integration of statistics guided optimization (SGO) principles with DNN-based architecture. Specifically, the parameters in the DDCCANet are determined by solving a SML-based optimization problem in each convolutional layer independently instead of the backpropagation (BP) algorithm, generating a deep-level feature representation with high quality.\\\indent Unlike the typical DNNs-based networks, the DDCCANet does not utilize the BP algorithm and the parallel computing of GPU during the training process. In [7], the DDCCANet is accomplished based on the CPU operation. Compared with GPU-based parallel computing, CPU is far less efficient and time consuming. In view of this problem, this paper presents a GPU-based acceleration algorithm to implement the DDCCANet algorithm. As a result, not only is the performance of the DDCCANet model guaranteed, but also greatly shortens the calculation time, making DDCCANet more applicable in real tasks. To verify the effectiveness of the proposed acceleration algorithm, we conduct experiments on three data sets with different scales: the ORL dataset, ETH-80 dataset and Caltech 256 dataset. The experiment results show the superiority of the proposed accelerated algorithm on the given tasks.\\\indent This work is organized as follows: the DDCCANet is briefly reviewed in Section II.  The proposed GPU-based accelerated algorithm is presented in Section III. Experimental results and analysis are conducted and analyzed in Section IV. Conclusions are given in Section V.
\section{The DDCCANet Model}
The proposed DDCCANet consists of three types of layers: DDCCA filters, pooling layer and information quality representation.
\subsection{DDCCA Filters}
Suppose we have a set of images $I=[I_{1},I_{2},\cdot \cdot \cdot I_{M}]$, where $M$ is the number of samples. Then, two views are generated from $I$, which are written as $I^{1}=[I_{1}^{1},I_{2}^{1},\cdot \cdot \cdot I_{M}^{1}]$ and $I^{2}=[I_{1}^{2},I_{2}^{2},\cdot \cdot \cdot I_{M}^{2}]$. Essentially, $I^{1}$ and $I^{2}$ can be considered as two 3-D tensors with a size of $M\times p\times q$, where $p$ and $q$ are the width and height of each image. Next, the two-view data sets are divided into patches with the size of $l_{1}\times l_{2}$ for each each sample pair. Then, all patches in $I^{1}$ and $I^{2}$ can be expressed as below
\begin{equation}
I^{d}=[I_{1}^{d*},I_{2}^{d*},\cdot \cdot \cdot I_{M}^{d*}]\in R^{l_{1}l_{2}\times Mpq}.
\end{equation}
Next, the discriminant canonical correlation analysis (DCCA) method [8] is used as optimization function to explore the discriminant feature representation between $I^{1}$ and $I^{2}$ in equation (2)
\begin{equation}
arg max \rho = w_{1}^{T}C_{I^{1}I^{2}}^{\sim }w_{2},
\end{equation}
subject to
\begin{equation}
w_{1}^{T}C_{I^{1}I^{1}}w_{1} = w_{2}^{T}C_{I^{2}I^{2}}w_{2}=\textup{I}.
\end{equation}
where $C_{I^{1}I^{2}}^{\sim }= {{C_{{\omega _{I^{1}I^{2}}}}}}- {{C_{{b_{I^{1}I^{2}}}}}}$, $C_{I^{1}I^{1}}=I^{1}{I^{1}}^{T}$ and $C_{I^{2}I^{2}}=I^{2}{I^{2}}^{T}$. In addition, ${{C_{{\omega _{I^{1}I^{2}}}}}}$ and ${{C_{{b_{I^{1}I^{2}}}}}}$ are within-class and between-class correlation matrices between $I^{1}$ and $I^{2}$. Then, it is necessary to find a pair of projected vectors $w_{1}$ and $w_{2}$ with the discriminant representations. After that, the convolutional filters. the $w_{1}$ and $w_{2}$ are reshaped with the following equations (4) and (5)
\begin{equation}
W_{g}^{1} = res_{l_{1},l_{2}}(w_{1,g})\in R^{l_{1}\times l_{2}},
\end{equation}
\begin{equation}
W_{g}^{2} = res_{l_{1},l_{2}}(w_{2,g})\in R^{l_{1}\times l_{2}},
\end{equation}
where $g=1,2,\cdot \cdot \cdot L_{i}$ and $L_{i}$ is the number of filters in the \emph{i}th layer. According to (4) and (5), the 2D convolution can be implemented as below:
\begin{equation}
	I_{d,k,g}^{out} = res_{l_{1},l_{2}}(I_{k}^{d*})\otimes W_{g}^{d}
\end{equation}
where $I_{d,k,g}^{out}$ is the generated feature map and $d$ denotes the $d$-th view.
\subsection{The pooling layer}
 Let $L_{i}$ be the number of filters in the \emph{i}th layer of each view. Then it leads to $L_{i}\times L_{i+1}$ feature maps in the \emph{i+1}th layer, resulting in the convergence due to the overfitting issue. To address this issue, a hashing pooling layer is utilized to reduce the size of the feature representation from the previous layers of each view. Assume the feature map is $I_{d,k,g}^{out}$ and the output feature map can be binarized with a Hashing transform as below
\begin{equation}
S\left \{ I_{d,k,g}^{out}\otimes W_{l}^{d} \right \},
\end{equation}
where $l=1,2,\cdot \cdot \cdot L_{i+1}$, and
\begin{equation}
S(x)=\left\{\begin{matrix}
	1 & (x>0),\\
	0 & other.
\end{matrix}\right.
\end{equation}
Since there are $L_{i+1}$ filter corresponding to each output from the previous layer, the pooling operator can be operated by (9)
\begin{equation}
Q_{d,k,g}=\sum_{l=1}^{L_{i+1}}2^{l-1}S\left \{ I_{d,k,g}^{out}\otimes W_{l}^{d} \right \}
\end{equation}
\subsection{The information quality-based fully connected layer}
In the DDCCANet model, an information quality (IQ)-based fully connected layer is constructed. The pooling result $Q_{d,k,g}$ from previous subsection is partitioned into A blocks, then IQ is utilized to produce the final deep-level feature representation. The definition of IQ is written in (10)
\begin{equation}
H(p(t)) = -log(p(t)).
\end{equation}
Finally, the generated deep-level feature representation of the k-th sample in the d-th view is expressed in (11)
\begin{equation}
o_{k,d}=[H(Q_{d,k,1})\cdot \cdot \cdot H(Q_{d,k,L_{i}})]^{T}\in R^{(2^{L(i+1)L_{i}A})}.
\end{equation}
Then, the two views-based deep-level feature representation is concatenated as follows
\begin{equation}
o_{k}=[o_{k,1},o_{k,2}].
\end{equation}
In summary, the proposed DDCCANet architecture is drawn in Figure. 1.
\begin{figure*}[t]
\centering
\includegraphics[height=2.4in,width=7.0in]{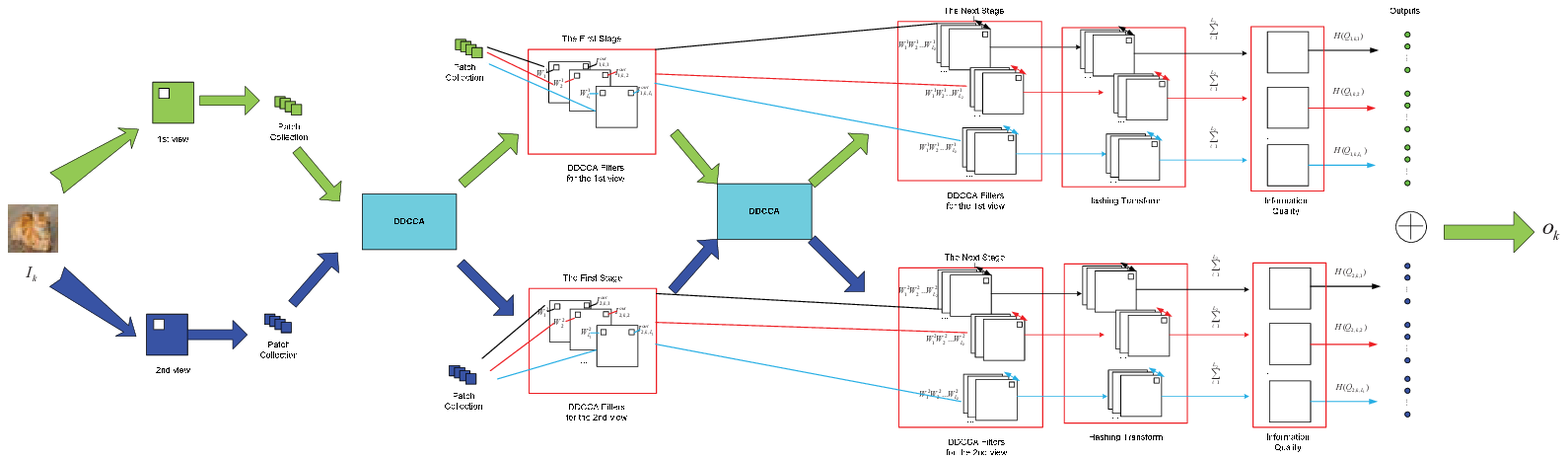}\\ Figure. 1 The diagram of the DDCCANet\\
\end{figure*}\\
\section{GPU-accelerated algorithm}
Distinct from the typical DNNs-based architectures, the BP algorithm is not utilized to train the DDCCANet model. In [7], the DDCCANet model was operated on the CPU platform. Although it has achieved impressive results, its implementation does not show clear advantage over state-of-the-art DNN models in terms of computational time. In order to handle this issue, a GPU acceleration algorithm is designed to improve implementation efficiency and computing time. The accelerated algorithm is given as follows.\\\indent First of all, all samples in a given dataset are splitted into several different batches for computing on GPU. Then in the DDCCANet driven multi-view convolutional layer, it needs to obtain the related parameters in convolutional filters (e.g., $C_{I^{1}I^{2}}^{\sim }$, $C_{I^{1}I^{2}}$, $C_{I^{1}I^{1}}$ and $C_{I^{2}I^{2}}$). Since $C_{I^{1}I^{1}}$, $C_{I^{2}I^{2}}$, and $C_{I^{1}I^{2}}^{\sim }$ are calculated by utilizing similar mathematical formula, an example on the calculation of $C_{I^{1}I^{1}}$ is given as follows.\\\indent In the two views data sets $I^{1}$ and $I^{2}$, all samples are splitted into $K$ batches, which are written as $batch\_I^{1,j}=B_{I^{1,j}}$ and $batch\_I^{2,j}=B_{I^{2,j}}$ ($j=1,2,\cdot \cdot \cdot K$). Then, $I^{1}$ and $I^{2}$ satisfies the relations in (13) and (14),\\
\begin{equation}
	I^{1}=\left [ B_{I^{1,1}},B_{I^{1,2}},...,B_{I^{1,K}} \right ]
\end{equation}

\begin{equation}
	I^{2}=\left [ B_{I^{2,1}},B_{I^{2,2}},...,B_{I^{2,K}} \right ]
\end{equation}

The parameter $C_{I^{1}I^{1}}$ is calculated by the product of the patches matrix and its transposition as shown in (15).\\
\begin{equation}
\begin{aligned}
	I^{1}\left (I^{1}  \right )^{T}&=[B_{I^{1,1}},B_{I^{1,2}},...,B_{I^{1,K}}]\begin{bmatrix}
		B_{I^{1,1}}^{T}\\
		B_{I^{1,1}}^{T}\\
		\vdots \\
		B_{I^{1,1}}^{T}
	\end{bmatrix}\\
	&= B_{I^{1,1}}B_{I^{1,1}}^{T}+B_{I^{1,2}}B_{I^{1,2}}^{T}+\cdots +B_{I^{1,K}}B_{I^{1,K}}^{T}
\end{aligned}
\end{equation}
According to equation (15), we can implement the DDCCANet model by the parallel computing of GPU. In the real implementation, we split samples into batched sets. The $K$ patches are obtained with the function GetPatches() for each batch. Similarly, for the parameter $C_{I^{1}I^{2}}$, we loop all classes, and all samples corresponding to each class to calculate $I^{1}$ and $I^{2}$. Then eachClass\_$C_{I^{1}I^{2}}$ is accumulated to calculate $C_{I^{1}I^{2}}$. Moreover, the parameter $C_{I^{1}I^{2}}^{\sim }$ is accomplished by utilizing the similar calculating procedure.\\\indent Finally, the convolutional filters $W^1$ and $W^2$ can be obtained with the DCCA method. With the convolutional filters ${W}^1$ and ${W}^2$, the convolutional results are calculated in batches. Then the feature representation $o$ is obtained after pooling and fully connected layer operation. The pseudo code is shown as follows: \\

\begin{figure*}[t]
	\centering
	\includegraphics[height=3in,width=4.5in]{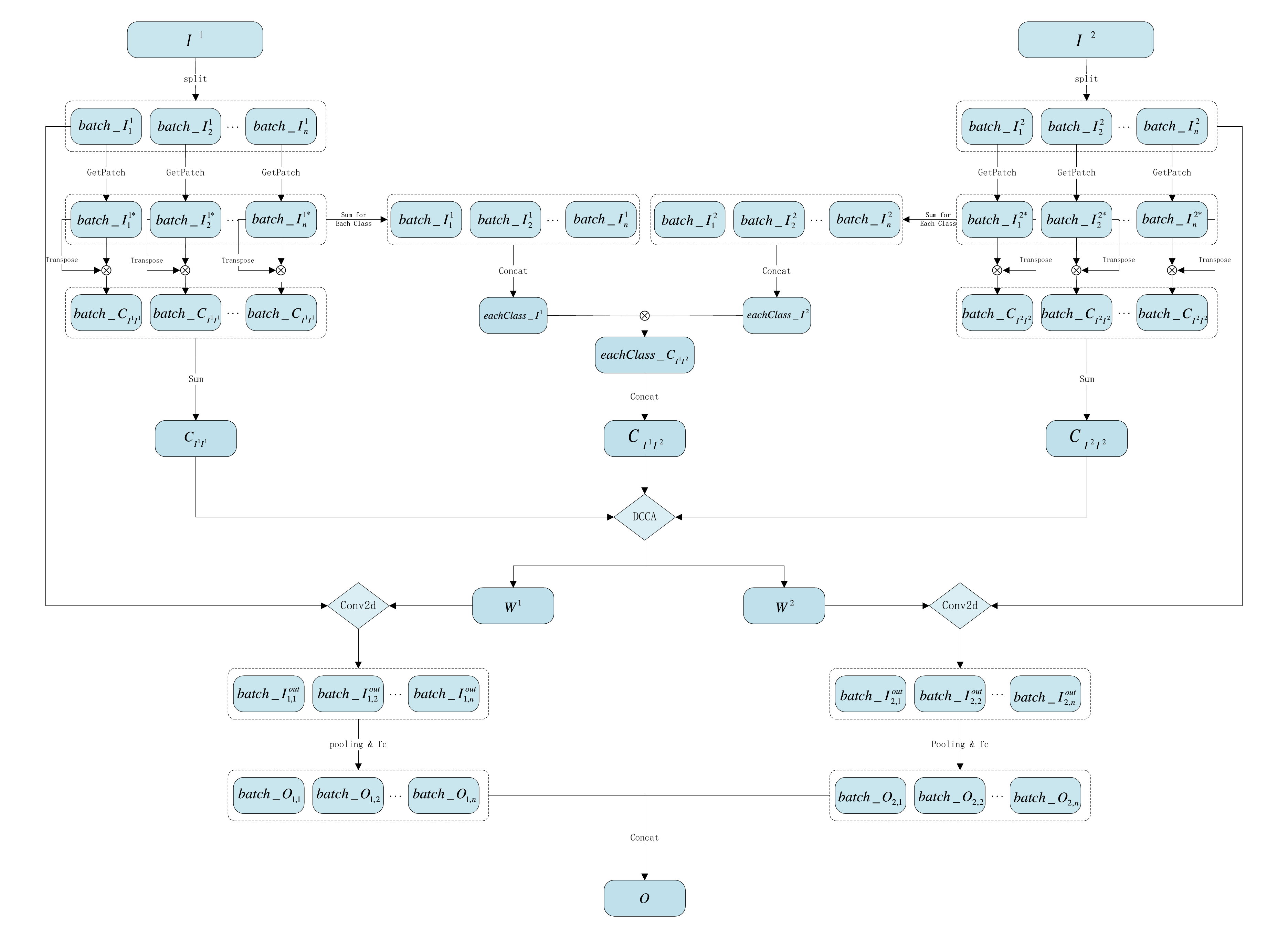}\\ Figure. 2 The diagram of the proposed GPU-based accelerated algorithm\\
\end{figure*}

~\\
for i in range(batches):

\quad batch\_$I_{i}^{1*}$=GetPatches($batch\_I_{i}^{1}$)

\quad batch\_$I_{i}^{2*}$=GetPatches($batch\_I_{i}^{2}$)

\quad batch\_$C_{I^{1}I^{1}}$=batch\_$I_{i}^{1*}$ $\times$ batch\_$I_{i}^{1*}$.t()

\quad batch\_$C_{I^{2}I^{2}}$=batch\_$I_{i}^{2*}$ $\times$ batch\_$I_{i}^{2*}$.t()

\quad $C_{I^{1}I^{1}}$+=batch\_$C_{I^{1}I^{1}}$

\quad $C_{I^{2}I^{2}}$+=batch\_$C_{I^{2}I^{2}}$

~\\

for j in range(classes)

\quad for i in range(batches):

\quad \quad batch\_$I_{i}^{1*}$=GetPatches($batch\_I_{i}^{1}$)

\quad \quad batch\_$I_{i}^{2*}$=GetPatches($batch\_I_{i}^{2}$)

\quad \quad batch\_$I_{i}^{1}$=Sum(batch\_$I_{i}^{1*}$(class==j))

\quad \quad batch\_$I_{i}^{2}$=Sum(batch\_$I_{i}^{2*}$(class==j))

\quad \quad eachClass\_$I^1$=Concatenate(eachClass\_$I^1$,$batch\_I_{i}^{1}$)

\quad \quad eachClass\_$I^2$=Concatenate(eachClass\_$I^2$,$batch\_I_{i}^{2}$)

\quad eachClass\_$C_{I^{1}I^{2}}$=eachClass\_$I^1 \times eachClass\_I^2$

\quad $C_{I^{1}I^{2}}$+=eachClass\_$C_{I^{1}I^{2}}$

~\\

$W^1$, $W^2$ = DCCA($C_{I^{1}I^{1}}$, $C_{I^{2}I^{2}}$, $C_{I^{1}I^{2}}$)

~\\

for i in range(batches):

\quad batch\_$I_{1,i}^{out}$=Conv2d($batch\_I_{i}^{1}$,$W^1$)

\quad batch\_$I_{2,i}^{out}$=Conv2d($batch\_I_{i}^{2}$,$W^2$)

\quad $I_{1}^{out}$=Concatenate($I_{1,i}^{out}$,batch\_$I_{1,i}^{out}$)

\quad $I_{2}^{out}$=Concatenate($I_{2,i}^{out}$,batch\_$I_{2,i}^{out}$)

~\\

for i in range(batches):

\quad batch\_$Q_{1,i}$=pooling(batch\_$I_{1,i}^{out}$)

\quad batch\_$Q_{2,i}$=pooling(batch\_$I_{2,i}^{out}$)

\quad batch\_$o_{1,i}$=fc(batch\_$Q_{1,i}$)

\quad batch\_$o_{2,i}$=fc(batch\_$Q_{2,i}$)

\quad batch\_$o_{i}$=Concatenate(batch\_$o_{1,i}$, batch\_$o_{2,i}$)

\quad $o$=Concatenate($o$, batch\_$o_{i}$)\\

According to the above description and analysis, the diagram of the proposed GPU-based accelerated algorithm is depicted in Figure. 2.\\

\section{Experiment and Performance Evaluation}
In this section, we conduct experiments on three database with different scales, including the ORL face database, ETH-80 database and Calthe256 database. During the experiments, a Nvidia RTX 3060 GPU with 12GB video memory is used for the implementation of the proposed GPU-accelerated algorithm. To balance performance and the GPU memory consumption, the DDCCANet with only two convolution layers is implemented on the GPU. In addition, all experiments are conducted 10 times and the average results are reported.
\subsection{The ORL Database}
The ORL database is a small scale data set and there are 40 subjects with 10 different images, leading to 400 samples in total. All the samples are captured under various environment conditions, such as facial expressions, posing, illumination, etc. In the experiments, five images of each subject are chosen for training while the remaining samples are utilized for testing. The local binary patterns (LBP) map and original image are utilized as the multi-view data sets. According to the DDCCANet model, two multi-view convolution layers are performed on the dataset, and the number of convolutional filters in the two layers is 8. Besides, the number of samples in each batch from two views is set to 128. Compared with the performance on CPU platform, the average performance on GPU is shown in Table 1. Comparison with state-of-the-art (SOTA) algorithms is reported in Table 2. It's shown that the GPU-accelerate DDCCANet with shallow layers (L=2) achieved the advanced performance.

\begin{center}
	\small
	\resizebox{\columnwidth}{!}{%
	\begin{tabular}{|c|c|c|c|l|}
		\hline
		\multicolumn{1}{|l|}{Methods}                                     & \multicolumn{1}{l|}{Layers} & \multicolumn{1}{l|}{Filters} & \multicolumn{1}{l|}{Accuracy} & Running Time \\ \hline
		\begin{tabular}[c]{@{}c@{}}DDCCANet \\ on CPU\end{tabular}        & 2                           &$L_1=L_2= 8$                            & 95.45\%                         & 500 seconds \\ \hline
		\begin{tabular}[c]{@{}c@{}}DDCCANet\\ GPU-accelerate\end{tabular} & 2                          &$L_1=L_2= 8$                           & 95.50\%                          & 20 seconds  \\ \hline
	\end{tabular}%
}
\end{center}
Table 1: The accuracy(\%) and running time on ORL dataset.\\
~\\
~\\
\vspace*{-10pt}
\begin{center}
	\begin{tabular}{|c|c|}
		\hline
		Methods                  & \multicolumn{1}{l|}{Performance} \\ \hline
		\textbf{GPU-accelerated DDCCANet} & \textbf{95.50\%}                          \\ \hline
		AOS+VGG {[}9{]}          & 93.62\%                          \\ \hline
		CDPL {[}10{]}            & 95.42\%                          \\ \hline
		SOLDE-TR {[}11{]}        & 95.03\%                          \\ \hline
		GDLMPP {[}12{]}          & 94.50\%                          \\ \hline
		CNN {[}13{]}             & 95.00\%                          \\ \hline
	\end{tabular}
\end{center}
Table 2: The performance with different algorithms on the ORL database.

\subsection{ ETH-80 Dataset and Results}
The ETH-80 database contains 3280 color RGB object samples. For each class, 410 images are selected for 10 different objects and each object is presented by 41 different viewpoint images. 1000 samples from this dataset are randomly choose as training data, and the remaining images are reserved for testing.  In the experiment, the raw data (R and G sub-channel images) are adopted as the two multi-view data sets. Based on the optimal performance, the DDCCANet consists of two multi-view convolution layers, and the number of convolutional filters of the two convolutional layers is 8. In the experiments, the number of samples in each batch is set to 128 and results are tabulated in Table 3. Furthermore, we conducted experiments with other SOTA algorithms and the results are given in Table 4.
\begin{center}
	\resizebox{\columnwidth}{!}{%
	\begin{tabular}{|c|c|c|c|l|}
		\hline
		\multicolumn{1}{|l|}{Methods}                                     & \multicolumn{1}{l|}{Layers} & \multicolumn{1}{l|}{Filters} & \multicolumn{1}{l|}{Accuracy} & Running Time \\ \hline
		\begin{tabular}[c]{@{}c@{}}ODMTCNet \\ on CPU\end{tabular}        & 2                           &$L_1=L_2= 8$                           & 92.24\%                         & 5400 seconds   \\ \hline
		\begin{tabular}[c]{@{}c@{}}ODMTCNet\\ GPU-accelerate\end{tabular} & 2                           &$L_1=L_2= 8$                           & 92.80\%                         & 180 seconds    \\ \hline
	\end{tabular}%
}
\end{center}
Table 3: The accuracy(\%) and running time on the ETH-80 dataset.\\
~\\
\vspace*{-10pt}
\begin{center}
	\begin{tabular}{|c|c|}
		\hline
		Methods                  & \multicolumn{1}{l|}{Performance} \\ \hline
		\textbf{GPU-accelerated DDCCANet} & \textbf{92.80\%}                          \\ \hline
		CCANet {[}5{]}           & 91.45\%                          \\ \hline
		PCANet {[}14{]}          & 91.28\%                          \\ \hline
		PML {[}15{]}             & 89.00\%                          \\ \hline
		MFD {[}16{]}             & 86.91\%                          \\ \hline
		ALP-TMR {[}17{]}         & 84.86\%                          \\ \hline
		CERML {[}18{]}           & 85.00\%                          \\ \hline
	\end{tabular}
\end{center}
Table 4: The performance with different algorithms on the ETH-80 database.\\

\subsection{ Caltech 256 Database}
The Caltech 256 database is a comparatively large-sized data set, which contains 22850 samples in 257 classes. In our experiments, 30 samples in each class are randomly chosen as training data, the remaining samples in each class are used for testing. During the experiment, the VGG-19 model is employed to extract DNN-based features. Specifically, two fully connected layers fc6 and fc7 are used to generate the two multi-view data sets. Corresponding to the best accuracy with DDCCANet, there are two convolutional layer are adopted on the Caltech 256 dataset, and the number of convolutional filters in the two convolutional layer is 8. In the real implementation, the number of samples in each batch is set to 128. The experimental settings and results are tabulated in Table 5. In order to demonstrate the effectiveness of the DDCCANet model, more comparison is conducted with SOTA methods and the results are tabulated in Table 6.

\begin{center}
	\resizebox{\columnwidth}{!}{%
	\begin{tabular}{|c|c|c|c|l|}
		\hline
		Methods                                                           & \multicolumn{1}{l|}{Layers} & \multicolumn{1}{l|}{Filters} & \multicolumn{1}{l|}{Accuracy} & Running Time \\ \hline
		\begin{tabular}[c]{@{}c@{}}ODMTCNet \\ on CPU\end{tabular}        & 2                           &$L_1=L_2= 8$                             & 86.27\%                         & $2.6\times 10^{5}$ seconds    \\ \hline
		\begin{tabular}[c]{@{}c@{}}ODMTCNet\\ GPU-accelerate\end{tabular} & 2                           &$L_1=L_2= 8$                             & 85.80\%                         & $1.8\times 10^{4}$ seconds     \\ \hline
	\end{tabular}%
}
\end{center}
Table 5: The accuracy(\%) and running time on Caltech 256 dataset.\\
~\\
~\\
\vspace*{-10pt}
\begin{center}
	\begin{tabular}{|c|c|}
		\hline
		Methods                             & \multicolumn{1}{l|}{Performance} \\ \hline
		\textbf{GPU-accelerated DDCCANet}   & \textbf{85.80\%}                 \\ \hline
		MVLS {[}19{]}                       & 84.23\%                          \\ \hline
		NR {[}20{]}                         & 84.40\%                          \\ \hline
		$L^{2}$-SP {[}21{]}                 & 84.90\%                          \\ \hline
		All+Wi-HSNN {[}22{]}                & 85.70\%                          \\ \hline
		T-ResNet+T-Inception+T-VGG {[}23{]} & 82.70\%                          \\ \hline
		Hybrid1365-VGG {[}24{]}             & 76.04\%                          \\ \hline
		EProCRC {[}25{]}                    & 84.30\%                          \\ \hline
		Hierarchical Networks {[}26{]}      & 84.10\%                          \\ \hline
		PtR {[}27{]}                        & 84.50\%                          \\ \hline
		TransTailor {[}28{]}                & 85.30\%                          \\ \hline
	\end{tabular}
\end{center}
Table 6: The performance with different algorithms on the Caltech 256.\\
\vspace*{-10pt}
\section{Conclusion}
 In this work, we propose a GPU-based accelerated algorithm for the DDCCANet Model. Different from the CPU-based algorithm, the proposed algorithm is able to substantially improve the implementing efficiency and computing time by utilizing the parallel computing power of GPU. The experimental results on three datasets of different scales validate the superiority of the proposed algorithm on given examples.





%\bibliographystyle{IEEEtran}

\end{document}